\documentclass[pre,twocolumn,showpacs,floatfix]{revtex4}

\usepackage{amsmath}
\usepackage{amssymb}
\usepackage{graphicx}
\begin{document}

\title{Origin of the anomalous heat current in collisional granular fluids}

\author{D. Candela}
\affiliation{Physics Department, University of Massachusetts, 
Amherst, MA 01003}

\author{R. L. Walsworth}
\affiliation{Harvard-Smithsonian Center for Astrophysics, Cambridge, MA 02138}

\begin{abstract}
	We present a heuristic explanation for the anomalous (density-gradient-dependent) heat current in collisional granular fluids. 
	Inelastic grain collisions lead to highly non-equilibrium states which are characterized by large spatial gradients and/or temporal variations in the granular temperature.
	It is argued that the heat current in such non-equilibrium states is driven by the temperature gradient averaged over the typical collision time and/or mean free path.
	Due to the density dependence of the inelastic energy loss, the nonlocal averaging of the temperature gradient leads to an effective dependence of the heat current upon the density gradient.
\end{abstract}

\date{7 October 2005}
\maketitle

	When energy is supplied to a system of inelastic grains, a particularly simple \emph{collisional} state can occur in which only binary collisions between the grains are important.
	The collisional state occurs for hard grains at any density below jamming, and for more realistic soft grains at densities sufficiently low that enduring and multiparticle contacts are rare~\cite{rapidnote,campbell02}.
	In recent decades collisional granular fluids have served as a paradigm for simulations and for the application of statistical ideas to granular 
systems~\cite{haff83,jenkins83}.
	One perplexing phenomenon long predicted~\cite{lun84} for collisional granular fluids is a violation of Fourier's law, which states that the heat current is proportional to the temperature gradient and always flows from hot to cold.
	In contrast, the heat current predicted by kinetic theory~\cite{lun84,sela98,brey98,garzo99} for a granular fluid is \begin{equation}\label{q}
	\mathbf{q} = -\kappa \nabla T -\mu \nabla n
\end{equation}
where $T = m \langle v^2 \rangle /d$ is the granular temperature in 
$d$ space dimensions for particles of mass $m$ and mean squared 
velocity $\langle v^2 \rangle$, and $n$ is the grain number density. 
Here $\kappa$ is the granular thermal conductivity and $\mu$ is a new 
transport coefficient that is nonzero for inelastic systems. Due to 
the last term in Eq.~\ref{q}, the \emph{anomalous heat 
current}, it is possible when $\nabla n \neq 0$ for the total heat current to flow from cold to hot.
	Although counterintuitive the anomalous heat current does not violate the second law of thermodynamics, as fluidized granular media are highly non-equilibrium systems.

\begin{figure}
\includegraphics[width=8.6 cm]{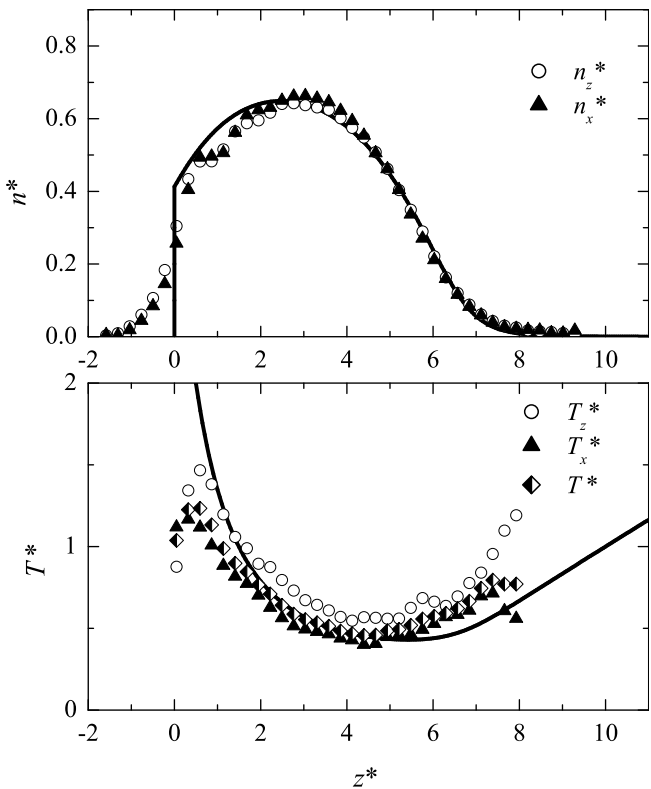}
\caption{\label{fig1}
	Measured granular density (top) and temperature (bottom) as functions of height for a dilute granular fluid that is vibrofluidized from below, from Ref.~\cite{huan04}.
	The symbols show NMR measurements for mustard seeds vibrated at $15g$ and 50~Hz, while the curves show a fit to the hydrodynamic theory of Ref.~\cite{garzo99}.
	Dimensionless units $z^* = z/\sigma$, $n^* = n\sigma^3$, $T^* = T/mg\sigma$ are used.
	The fitted restitution coefficient was $\alpha = 0.87$.
	The sample was confined by gravity to the bottom region of the sample chamber with a free upper surface, so there was no vibrational energy input from above.
	Nevertheless, the experimental data show a clear minimum in the granular temperature $T^*$ as a function of height $z^*$.
	At heights above the temperature minimum, the heat current (always upward) is from cold to hot.
}\end{figure}

	In the presence of gravity the anomalous heat current is manifested by a nonzero heat current at the height at which the temperature gradient is zero~\cite{soto99}.
	More strikingly, for a system with a free upper surface (confined by 
gravity and excited from below), the anomalous heat current can lead 
to a \emph{temperature inversion}: despite the strictly upward heat 
current, the temperature first falls as a function of height and then 
rises again in the vicinity of the free upper surface~\cite{ramirez03}.
	In a recent experiment using NMR to probe vibrofluidized mustard seeds, Huan et al.~\cite{huan04} observed such a temperature inversion and 
quantitatively fit the heat current (including the anomalous term) to 
theoretical predictions.
	 Fig.~\ref{fig1} shows example data from this study along with the hydrodynamic fit.
	A temperature inversion may also have been observed in much earlier experiments on vibrofluidized systems~\cite{clement91,warr95}, but the inversion was not associated at the time with the anomalous heat current.

	Although a density-gradient-dependent heat current has been 
demonstrated in numerical simulations of granular 
fluids~\cite{soto99,ramirez03}, and $\mu$ has been calculated using 
sophisticated statistical-physics 
methods~\cite{lun84,sela98,brey98,garzo99}, a simple heuristic explanation of this phenomenon has not been presented to date.
	In this paper we present such an explanation, tracing the anomalous heat current directly to the highly non-equilibrium nature of collisional granular fluids.
	Due to inelastic energy losses, such fluids necessarily have large time derivatives and/or spatial gradients of the granular temperature.
	As the temperature may vary significantly over a mean-free-path distance or grain collision time, the inherent \emph{nonlocality} of the grain collision process must be taken into account.
	By considering the consequences of this nolocality, we show that the density dependence of the inelastic energy loss leads to the anomalous heat current term.
	We argue that the total heat current is fundamentally driven by temperature gradients, not density gradients, and that the apparent dependence of the heat current upon density gradients arises from re-expressing nonlocal quantities in terms of local ones.

	The existing technical calculations~\cite{sela98,brey98,garzo99} of the anomalous transport coefficient $\mu$ are valid for wide ranges of the dimensionless density $n \sigma^3$ and the inelasticity parameter $1-\alpha$ (here $\sigma$ is the grain diameter and $\alpha$ is the normal restitution coefficient).
	To validate our explanation for the anomalous heat current, we show that it predicts a value for $\mu$ that agrees up to factors of order unity with these more rigorous calculations in the dilute, nearly elastic limit $n\sigma^3 \ll 1$, $(1-\alpha) \ll 1$.
	The aim of the present paper is not to reproduce precisely the earlier results, but rather to provide a simple, physically-motivated explanation for the last term in Eq.~\ref{q}.

	We begin with the commonly-used relaxation-time approximation for the evolution of the heat current:
 \begin{equation}\label{dqdt}
	\frac{\partial \mathbf{q}}{\partial t} =
		-\frac{b_q nT}{m} \nabla T - \nu_q \mathbf{q},
\end{equation}
where $\nu_q = b_\nu n\sigma^{d-1}(T/m)^{1/2}$ is the relaxation rate 
for heat currents, roughly equal to the grain collision frequency.
	Here and below, $b_q, b_\nu \ldots$ are dimensionless constants of order unity that depend only upon the 
dimension of space $d$; values are given in Table~\ref{abcdtab}. Note 
that $(T/m)^{1/2}$ is a typical particle speed and 
$\sigma^{d-1}$ is roughly the collision cross section.

\begin{table}\caption{\label{abcdtab}
	Dimensionless prefactors used in kinetic theory expressions 
in this paper, for dimensions $d=2$ and 3.  The $b_q$ values are 
calculated for a Maxwell-Boltzmann velocity distribution, expected to 
be a good approximation for weakly inelastic systems~\cite{brey99}.
	The $b_\nu$ and $b_\zeta$ values are computed from the 
results tablulated in Appendix~A of Ref.~\cite{brey01}.
	Finally, the $b_\lambda$ values are those required for the 
anomalous transport coefficient $\mu$ to be the same for cooling 
states and steady states according to the approximate calculation 
presented in this paper.
}\begin{ruledtabular}
\begin{tabular}{cccc}
Prefactor & Quantity  & $d=2$ & $d=3$\\
\hline
$b_q$ & Heat current & $4/3$ & $5/2$ \\
$b_\nu$ & Relaxation frequency & $2\sqrt{\pi}/3$ & $32\sqrt{\pi}/15$ \\ $b_\zeta$ & Cooling rate & $2\sqrt{\pi}$ & $8\sqrt{\pi}/3$ \\ $b_\lambda$ & Mean free path & $2$ & $\sqrt{5}$ \\ \end{tabular} \end{ruledtabular} \end{table}

	The first term on the RHS of Eq.~\ref{dqdt} gives the buildup of the heat current due to the free streaming of particles from nearby regions with differing temperatures, while the second term expresses the relaxation of $\mathbf{q}$ towards zero due to collisions, which randomize the particle velocities.
	In an \emph{elastic} system ($\alpha=1$) perturbed by weak temperature and density gradients, the temperature varies slowly in time due to energy conservation.
	Hence a steady state is reached with $\partial\mathbf{q}/\partial t \approx 0$, giving Fourier's law: $\mathbf{q} = -\kappa_0 \nabla T$ with $\kappa_0 = b_q nT/m\nu_q \sim (T/m)^{1/2}/\sigma^{d-1}$.

	Implicit in this derivation of Fourier's law are two related ``near-equilibrium'' assumptions: (i) dynamics affecting $n$ and $T$ occur very slowly relative to the heat current relaxation rate $\nu_q$, and (ii) spatial variations of  $n$ and $T$ occur on characteristic length scales much greater than the mean free path $\lambda_q \sim 1/n\sigma^{d-1}$.
	In an \emph{inelastic} system ($\alpha<1$) the near-equilibrium assumptions are not generally valid because the system continually loses energy, which causes temperature changes at rates comparable to $\nu_q$.
	If energy is supplied at the boundary of the system to maintain a steady state, this leads to density and temperature variations on length scales comparable to $\lambda_q$.
	To account for these non-equilibrium effects the derivation of Fourier's law must be generalized to be nonlocal in space and time.
	We show here how this leads to an effective dependence of the heat current upon $\nabla n$.

	As a first example, we consider a \emph{freely-cooling granular fluid} (i.e., no energy input) with small gradients in $T$ and $n$.
	Locally, one has $\partial T/\partial t = -\zeta T$ with the cooling rate $\zeta$ given~\cite{brey98} to leading order in the inelasticity $(1-\alpha)$ by:
\begin{equation}\label{zeta}
	\zeta = b_\zeta 
n\sigma^{d-1}(T/m)^{1/2}[(1-\alpha)+\mathcal{O}(1-\alpha)^2].
\end{equation}
	For typical values of the restitution coefficient $\alpha \approx 0.9$, the ratio of the cooling rate to the heat current relaxation rate, $\zeta/\nu_q = (b_\zeta/b_\nu)(1-\alpha)$, is a small but not infinitesimal quantity ($\approx 0.1$ in 3D and 0.3 in 2D). 
	Therefore $T$ changes non-negligibly during a typical grain collision time $1/\nu_q$, which violates the first near-equilibrium assumption above.

	We use the following simple argument to estimate the heat current $\mathbf{q}(t)$ in this situation.
	According to Eq.~\ref{dqdt} $\mathbf{q}(t)$ continuously relaxes towards the Fourier-law value $-\kappa_0 \nabla T$, which however is time-dependent.
	Thus, $\mathbf{q}(t)$ lags behind the Fourier-law value by approximately one relaxation time $1/\nu_q$:
\begin{eqnarray}\label{qt}
	\mathbf{q}(t) & \approx & -\kappa_0 \nabla T + 
(1/\nu_q)\partial(\kappa_0 \nabla T)/\partial t.
\end{eqnarray}
	Equation \ref{qt} expresses the idea that the gradient that drives the heat current is \emph{nonlocal in time}:  The heat current at time $t$ reflects the weighted average of $-\kappa_0 \nabla T$ over earlier times, with an typical lag of one collision time $1/\nu_q$.
	The second term in Eq.~\ref{qt} is readily calculated,
\begin{eqnarray}\label{dgtdt}
	\partial(\kappa_0 \nabla T)/\partial t & = & 
\frac{\partial\kappa_0}{\partial t}\nabla T + \kappa_0 \nabla(-\zeta 
T) \nonumber\\
		& = & -\frac{3}{2}\kappa_0 \zeta \nabla T
- \kappa_0 T \left[\frac{\partial\zeta}{\partial T}\nabla T
			+\frac{\partial\zeta}{\partial n}\nabla n 
\right] \nonumber\\
		& = & -2 \kappa_0 \zeta \nabla T - \frac{\kappa_0 
\zeta T}{n} \nabla n
\end{eqnarray}
where $\zeta \propto nT^{1/2}$ and $\kappa_0 \propto T^{1/2}$ have 
been used to simplify this equation.
	Thus:
\begin{equation}\label{q2}
	\mathbf{q} \approx -\left(1+2\frac{\zeta}{\nu_q}\right)\kappa_0\nabla T
			- \frac{\kappa_0 \zeta T}{\nu_q n}\nabla n.
\end{equation}
	From Eq.~\ref{q2} we can read off the anomalous transport coefficient $\mu = \kappa_0 \zeta T/\nu_q n$, proportional to the semi-small parameter $\zeta/\nu_q$.
	The ordinary thermal conductivity also acquires a correction proportional to the same parameter.
	Both of these results are close to the results of more sophisticated calculations~\cite{lun84,sela98,brey98,garzo99}, for example the value of $\mu$ is $5/4$ times the result calculated in Ref.~\cite{brey98}.

	The origin of the heat current's effective density-gradient-dependence is the dependence of the cooling rate on density ($\partial\zeta/\partial n \neq 0$ in Eq.~\ref{dgtdt}).
	For example, if at time $t=0$ there is a density gradient but no temperature gradient, then at earlier times the denser regions must have been hotter since they are cooling faster.
	Therefore, averaging $\kappa_0 \nabla T$ over times $t<0$ yields a non-zero heat current in the absence of a temperature gradient at $t=0$.

		As a second example we consider a \emph{boundary-driven steady-state} --- e.g., a vertically-vibrated granular medium such as that represented in the data of Fig.~\ref{fig1} and described in Ref.~\cite{huan04}.
	Due to inelastic energy losses and gravity, a boundary-driven steady-state necessarily has strong spatial variations of $n$ and $T$, in violation of the second near-equilibrium assumption listed above.

	A rigorous treatment of transport in the presence of strong gradients 
is complex, but the folowing simple argument should give a reasonable approximation.
	The typical distance over which a particle drifts before having its velocity randomized by a collision is the mean free path $\lambda_q = b_\lambda(T/m)^{1/2}/\nu_q \sim 1/n\sigma^{d-1}$, where $b_\lambda$ is another dimensionless constant of order one.
	Therefore we should replace $\nabla T$ in Eq.~\ref{dqdt} by $\langle \nabla T \rangle_\lambda$, which denotes the spatial average of $\nabla T$ over the mean free path $\lambda_q$.
	Thus, in the steady state the temperature gradient is effectively \emph{nonlocal in space}, just as it is effectively nonlocal in time for the cooling state.

\begin{figure}
\includegraphics[width=8.6 cm]{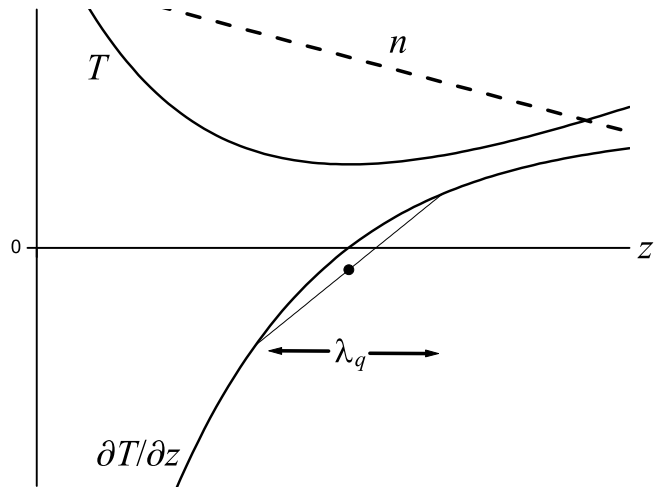}
\caption{\label{fig2}
	Schematic illustration of the mechanism for anomalous heat 
currents in a steady-state situation similar to that shown in 
Fig.~\ref{fig1}.
	At the height $z$ of the temperature minimum, the gradient 
$\partial T/\partial z$ has non-zero curvature due to the density 
gradient and the density dependence of the inelastic power loss.
	As a result of the curvature, the average of the temperature 
gradient over the mean free path $\lambda_q$ is non-zero (filled dot 
on graph).
	This non-zero effective gradient in turn drives a heat 
current even though $\partial T/\partial z$ is zero at this height.
	(Although it might appear from this diagram that an anomalous 
heat current can only occur over a small range of $z$ near the 
minimum, this is not the case due to the increase of $\lambda_q$ with 
$z$.)
}\end{figure}

	In the steady state $\langle \nabla T \rangle_\lambda$ differs from $\nabla T$ by an amount proportional to the curvature of 
$\nabla T$ times $\lambda_{q}^2$, as shown schematically in Fig.~\ref{fig2}.
	To simplify the calculation, we assume that $T$ and $n$ depend only upon the $z$ coordinate (typically in the direction opposite gravity).
Then
\begin{eqnarray}\label{gtav}
	\langle \partial T/\partial z \rangle_\lambda & = &
	\frac{1}{2\lambda_q} \int_{-\lambda_q}^{\lambda_q}
	\frac{\partial T}{\partial z}(z+u)du \nonumber\\
		& \approx & \partial T/\partial z + (\lambda_q^2/6) 
\partial T^3/{\partial z^3}.
\end{eqnarray}
	To lowest order in $\zeta/\nu_q$ we have $\partial T/\partial z = 
-q_z/\kappa_0$, where $q_z$ is the $z$-component of the heat current 
$\mathbf{q}$.
	Hence
\begin{equation}\label{d3t}
\partial T^3/\partial z^3 \approx -\kappa_0^{-1}\partial^2 q_z /\partial z^2. \end{equation}
	This omits corrections proportional to derivatives of $\kappa_0$, which considerably complicate the calculation but do not affect the leading-order estimate for $\mu$.

	To complete the argument we note that in a steady state the energy loss rate per unit volume due to inelastic collisions, $P_c$, is balanced by the negative divergence of the heat current: $P_c = -\nabla \cdot \mathbf{q} = -\partial q_z/\partial z$.
	The energy loss rate is proportional to the cooling rate 
$\zeta$: $P_c = c\zeta T = d n\zeta T/2$ where $c = dn/2$ is the specific heat~\cite{heatnote}.
	Using Eqs.~\ref{gtav} and \ref{d3t} we have
\begin{eqnarray}
	\langle \partial T/\partial z \rangle_\lambda  -
		\partial T/\partial z & \approx & -(\lambda_q^2/6\kappa_0) \partial^2 q_z/\partial z^2 \nonumber\\
	& = & (\lambda_q^2 d/12\kappa_0)\partial(n\zeta T)/\partial z \nonumber\\
	& = & \frac{\lambda_q^2 d\zeta n}{6\kappa_0}\left[ 
\frac{3}{4}\frac{\partial T}{\partial z} + \frac{T}{n}\frac{\partial 
n}{\partial z} \right]
\end{eqnarray}
where $\zeta \propto nT^{1/2}$ has again been used.
	Thus we find for the steady-state heat current:
\begin{eqnarray}\label{qz}
	q_z & = & -\kappa_0 \langle \partial T/\partial z 
\rangle_\lambda \nonumber\\
	& = & -\kappa_0\left(1 + 
\frac{b_\lambda^2 d\zeta}{8b_q\nu_q}\right) \frac{\partial T}{\partial 
z} - \frac{b_\lambda^2 d\kappa_0 \zeta T}{6 b_q \nu_q n}\frac{\partial 
n}{\partial z},
\end{eqnarray}
giving $\mu = b_{\lambda}^2 d\kappa_0 \zeta T/6 b_q \nu_q n$ for the anomalous heat transport coefficient in a boundary-driven steady-state.
	This value for $\mu$ in the steady state agrees with the cooling-state estimate of $\mu$ derived above for $b_\lambda = (6 b_q/d)^{1/2}$, which gives reasonable values for the dimensionless constant $b_\lambda$ (Table~\ref{abcdtab}).

	For the boundary-driven steady-state, the origin of the anomalous heat current is again the density dependence of the collisional energy loss ($P_c = c\zeta T \sim n^2$), which forces the heat current and hence the temperature gradient to have a large, non-zero second spatial derivative (curvature).
	The curvature of the temperature gradient over distances comparable to the mean free path in turn implies that the effective (nonlocal) temperature gradient $\langle \nabla T \rangle_\lambda$ is different from the local gradient, leading to an extra term in the heat current that is proportional to the density gradient (Fig.~\ref{fig2}).

	In summary, although the anomalous heat current is nominally proportional to the density gradient, an analysis of the underlying physical effects shows that the driving term is actually the temperature gradient.
	Due to the large time and/or space derivatives of the temperature that necessarily occur in dilute granular fluids, the driving temperature gradient must be averaged over a typical collision time $1/\nu_q$ and mean free path $\lambda_q$.
	Thus the effective gradient is nonlocal in time and space; when expressed in terms of local gradients an apparent density-driven heat current appears.

	Finally, we emphasize that it is the strong space or time derivative of the temperature, rather than the inelasticity \emph{per se}, that is the source of the anomalous heat current.
	Therefore, similar heat currents should in principle be observable in elastic systems if suitably strong temperature gradients could be imposed.
	From this point of view, the role of inelasticity in a granular fluid is primarily to create a highly non-equilibrium state, and some novel physical properties of granular fluids (like the anomalous heat current) are due fundamentally to the large deviation from equilibrium rather than the inelastic energy loss.

	We thank A. Santos for useful communications including a derivation of $\mu$ in the cooling state which motivated the heuristic argument given in the first part of this paper.
We also thank W. M. Mullin for calculation of the $b_q$ values in Table~\ref{abcdtab}.
	This work was supported by NSF Grant No. CTS-0310006.

\bibliography{granflow,mu}

\begin{thebibliography}{16}
\expandafter\ifx\csname natexlab\endcsname\relax\def\natexlab#1{#1}\fi
\expandafter\ifx\csname bibnamefont\endcsname\relax
  \def\bibnamefont#1{#1}\fi
\expandafter\ifx\csname bibfnamefont\endcsname\relax
  \def\bibfnamefont#1{#1}\fi
\expandafter\ifx\csname citenamefont\endcsname\relax
  \def\citenamefont#1{#1}\fi
\expandafter\ifx\csname url\endcsname\relax
  \def\url#1{\texttt{#1}}\fi
\expandafter\ifx\csname urlprefix\endcsname\relax\def\urlprefix{URL }\fi
\providecommand{\bibinfo}[2]{#2}
\providecommand{\eprint}[2][]{\url{#2}}

\bibitem[{rap()}]{rapidnote}
\bibinfo{note}{Collisional states are often called \emph{rapid granular flows}.
  They can occur, however, in the absence of bulk flow as in the situations
  analyzed here.}

\bibitem[{\citenamefont{Campbell}(2002)}]{campbell02}
\bibinfo{author}{\bibfnamefont{C.~S.} \bibnamefont{Campbell}},
  \bibinfo{journal}{J. Fluid Mech.} \textbf{\bibinfo{volume}{465}},
  \bibinfo{pages}{261} (\bibinfo{year}{2002}).

\bibitem[{\citenamefont{Haff}(1983)}]{haff83}
\bibinfo{author}{\bibfnamefont{P.~K.} \bibnamefont{Haff}}, \bibinfo{journal}{J.
  Fluid Mech.} \textbf{\bibinfo{volume}{134}}, \bibinfo{pages}{401}
  (\bibinfo{year}{1983}).

\bibitem[{\citenamefont{Jenkins and Savage}(1983)}]{jenkins83}
\bibinfo{author}{\bibfnamefont{J.~T.} \bibnamefont{Jenkins}} \bibnamefont{and}
  \bibinfo{author}{\bibfnamefont{S.~B.} \bibnamefont{Savage}},
  \bibinfo{journal}{J. Fluid Mech.} \textbf{\bibinfo{volume}{130}},
  \bibinfo{pages}{187} (\bibinfo{year}{1983}).

\bibitem[{\citenamefont{Lun et~al.}(1984)\citenamefont{Lun, Savage, Jeffrey,
  and Chepurniy}}]{lun84}
\bibinfo{author}{\bibfnamefont{C.~K.~K.} \bibnamefont{Lun}},
  \bibinfo{author}{\bibfnamefont{S.~B.} \bibnamefont{Savage}},
  \bibinfo{author}{\bibfnamefont{D.~J.} \bibnamefont{Jeffrey}},
  \bibnamefont{and}
  \bibinfo{author}{\bibfnamefont{N.}~\bibnamefont{Chepurniy}},
  \bibinfo{journal}{J. Fluid Mech.} \textbf{\bibinfo{volume}{140}},
  \bibinfo{pages}{223} (\bibinfo{year}{1984}).

\bibitem[{\citenamefont{Sela and Goldhirsch}(1998)}]{sela98}
\bibinfo{author}{\bibfnamefont{N.}~\bibnamefont{Sela}} \bibnamefont{and}
  \bibinfo{author}{\bibfnamefont{I.}~\bibnamefont{Goldhirsch}},
  \bibinfo{journal}{J. Fluid Mech.} \textbf{\bibinfo{volume}{361}},
  \bibinfo{pages}{41} (\bibinfo{year}{1998}).

\bibitem[{\citenamefont{Garz{\'{o}} and Dufty}(1999)}]{garzo99}
\bibinfo{author}{\bibfnamefont{V.}~\bibnamefont{Garz{\'{o}}}} \bibnamefont{and}
  \bibinfo{author}{\bibfnamefont{J.~W.} \bibnamefont{Dufty}},
  \bibinfo{journal}{Phys. Rev. E} \textbf{\bibinfo{volume}{59}},
  \bibinfo{pages}{5895} (\bibinfo{year}{1999}).

\bibitem[{\citenamefont{Brey et~al.}(1998)\citenamefont{Brey, Dufty, Kim, and
  Santos}}]{brey98}
\bibinfo{author}{\bibfnamefont{J.~J.} \bibnamefont{Brey}},
  \bibinfo{author}{\bibfnamefont{J.~W.} \bibnamefont{Dufty}},
  \bibinfo{author}{\bibfnamefont{C.~S.} \bibnamefont{Kim}}, \bibnamefont{and}
  \bibinfo{author}{\bibfnamefont{A.}~\bibnamefont{Santos}},
  \bibinfo{journal}{Phys. Rev. E} \textbf{\bibinfo{volume}{58}},
  \bibinfo{pages}{4638} (\bibinfo{year}{1998}).

\bibitem[{\citenamefont{Soto et~al.}(1999)\citenamefont{Soto, Mareschal, and
  Risso}}]{soto99}
\bibinfo{author}{\bibfnamefont{R.}~\bibnamefont{Soto}},
  \bibinfo{author}{\bibfnamefont{M.}~\bibnamefont{Mareschal}},
  \bibnamefont{and} \bibinfo{author}{\bibfnamefont{D.}~\bibnamefont{Risso}},
  \bibinfo{journal}{Phys. Rev. Lett.} \textbf{\bibinfo{volume}{83}},
  \bibinfo{pages}{5003} (\bibinfo{year}{1999}).

\bibitem[{\citenamefont{Ram{\'{i}}rez and Soto}(2003)}]{ramirez03}
\bibinfo{author}{\bibfnamefont{R.}~\bibnamefont{Ram{\'{i}}rez}}
  \bibnamefont{and} \bibinfo{author}{\bibfnamefont{R.}~\bibnamefont{Soto}},
  \bibinfo{journal}{Physica A} \textbf{\bibinfo{volume}{322}},
  \bibinfo{pages}{73} (\bibinfo{year}{2003}).

\bibitem[{\citenamefont{Huan et~al.}(2004)\citenamefont{Huan, Yang, Candela,
  Mair, and Walsworth}}]{huan04}
\bibinfo{author}{\bibfnamefont{C.}~\bibnamefont{Huan}},
  \bibinfo{author}{\bibfnamefont{X.}~\bibnamefont{Yang}},
  \bibinfo{author}{\bibfnamefont{D.}~\bibnamefont{Candela}},
  \bibinfo{author}{\bibfnamefont{R.~W.} \bibnamefont{Mair}}, \bibnamefont{and}
  \bibinfo{author}{\bibfnamefont{R.~L.} \bibnamefont{Walsworth}},
  \bibinfo{journal}{Phys. Rev. E} \textbf{\bibinfo{volume}{69}},
  \bibinfo{pages}{041302} (\bibinfo{year}{2004}).

\bibitem[{\citenamefont{Cl{\'{e}}ment and Rajchenbach}(1991)}]{clement91}
\bibinfo{author}{\bibfnamefont{E.}~\bibnamefont{Cl{\'{e}}ment}}
  \bibnamefont{and}
  \bibinfo{author}{\bibfnamefont{J.}~\bibnamefont{Rajchenbach}},
  \bibinfo{journal}{Europhys. Lett.} \textbf{\bibinfo{volume}{16}},
  \bibinfo{pages}{133} (\bibinfo{year}{1991}).

\bibitem[{\citenamefont{Warr et~al.}(1995)\citenamefont{Warr, Huntley, and
  Jacques}}]{warr95}
\bibinfo{author}{\bibfnamefont{S.}~\bibnamefont{Warr}},
  \bibinfo{author}{\bibfnamefont{J.~M.} \bibnamefont{Huntley}},
  \bibnamefont{and} \bibinfo{author}{\bibfnamefont{G.~T.~H.}
  \bibnamefont{Jacques}}, \bibinfo{journal}{Phys. Rev. E}
  \textbf{\bibinfo{volume}{52}}, \bibinfo{pages}{5583} (\bibinfo{year}{1995}).

\bibitem[{\citenamefont{Brey et~al.}(1999)\citenamefont{Brey, Cubero, and
  Ruiz-Montero}}]{brey99}
\bibinfo{author}{\bibfnamefont{J.~J.} \bibnamefont{Brey}},
  \bibinfo{author}{\bibfnamefont{D.}~\bibnamefont{Cubero}}, \bibnamefont{and}
  \bibinfo{author}{\bibfnamefont{M.~J.} \bibnamefont{Ruiz-Montero}},
  \bibinfo{journal}{Phys. Rev. E} \textbf{\bibinfo{volume}{59}},
  \bibinfo{pages}{1256} (\bibinfo{year}{1999}).

\bibitem[{\citenamefont{Brey et~al.}(2001)\citenamefont{Brey, Ruiz-Montero, and
  Moreno}}]{brey01}
\bibinfo{author}{\bibfnamefont{J.~J.} \bibnamefont{Brey}},
  \bibinfo{author}{\bibfnamefont{M.~J.} \bibnamefont{Ruiz-Montero}},
  \bibnamefont{and} \bibinfo{author}{\bibfnamefont{F.}~\bibnamefont{Moreno}},
  \bibinfo{journal}{Phys. Rev. E} \textbf{\bibinfo{volume}{63}},
  \bibinfo{pages}{061305} (\bibinfo{year}{2001}).

\bibitem[{hea()}]{heatnote}
\bibinfo{note}{It might be appear that if grain rotations were coupled to
  translational motion, however weakly, the specific heat would increase from
  $dn/2$ to $(2d-1)n/2$ thus changing the value of $\mu$. However, if the power
  loss $P_c$ remained constant the cooling rate $\zeta$ would decrease by the
  same factor, leaving $\mu$ unchanged.}

\end{thebibliography}
\end{document}